\documentclass[useAMS,usenatbib]{mn2e}
\usepackage{color}
\usepackage{graphicx}

\def\be{\begin{equation}} 
\def\ee{\end{equation}}

\def\HI{\hbox{H~$\scriptstyle\rm I\ $}} 
\def\HII{\hbox{H~$\scriptstyle\rm II\ $}}

\def\gsim{\lower.5ex\hbox{\gtsima}} 
\def\lsim{\lower.5ex\hbox{\ltsima}} \def\gtsima{$\; \buildrel > \over 
\sim \;$} \def\ltsima{$\; \buildrel < \over \sim \;$} \def\prosima{$\; 
\buildrel \propto \over \sim \;$} \def\gsim{\lower.5ex\hbox{\gtsima}} 
\def\lsim{\lower.5ex\hbox{\ltsima}} 
\def\simgt{\lower.5ex\hbox{\gtsima}} 
\def\simlt{\lower.5ex\hbox{\ltsima}} 
\def\simpr{\lower.5ex\hbox{\prosima}} \def\la{\lsim} \def\ga{\gsim}

\def\gtsima{$\; \buildrel > \over \sim \;$} 
\def\ltsima{$\; \buildrel < \over \sim \;$} 
\def\gsim{\lower.5ex\hbox{\gtsima}} 
\def\lsim{\lower.5ex\hbox{\ltsima}} 
\def\simgt{\lower.5ex\hbox{\gtsima}} 
\def\simlt{\lower.5ex\hbox{\ltsima}} 
\def\simpr{\lower.5ex\hbox{\prosima}} 
\def\la{\lsim} 
\def\ga{\gsim}

\def\E3{{\cal E}_{\rm g}^{III}}

\title[Local Group progenitors: Ly$\alpha$ bright?]{Local Group progenitors: Lyman Alpha bright?}
\author[Dayal \& Libeskind]{Pratika Dayal$^{1}$,  \& Noam I Libeskind$^{1}$\\ 
$^{1}$ Leibniz-Institute for Astrophysics, Potsdam, An der Sternwarte 16, Potsdam, Germany, 14482 }

\begin{document} 
\date{} 
\pagerange{\pageref{firstpage}--\pageref{lastpage}} \pubyear{} 
\maketitle 

\label{firstpage} 
\begin{abstract}

We present a novel approach of identifying the Milky Way (MW) and Andromeda (M31) progenitors that could be visible as LAEs at $z \sim 6$: we couple a snapshot from the Constrained Local UniversE Simulations (\texttt{CLUES}) project, that successfully reproduces the MW and M31 galaxies situated in their correct environment, to a Lyman Alpha Emitter (LAE) model. Exploring intergalactic medium (IGM) ionization states ranging from an almost neutral to a fully ionized one, we find that including (excluding) the effects of clustered sources the first local group progenitor appears as a LAE for a neutral hydrogen fraction $\chi_{HI} =0.4$ ($\chi_{HI}= 0.1$). This number increases to 5 progenitors each of the MW and M31 being visible as LAEs for $\chi_{HI} = 10^{-5}$; the contribution from clustered sources is crucial in making many of the progenitors visible in the Ly$\alpha$, for all the ionization states considered. The stellar mass of the local group LAEs ranges between $10^{7.2-8} {\rm M_\odot}$, the dust mass is between $10^{4.6-5.1} {\rm M_\odot}$ and the color excess $E(B-V)=0.03-0.048$. We find that the number density of these LAEs are higher than that of general field LAEs (observed in cosmological volumes) by about two (one) orders of magnitude for $\chi_{HI}=10^{-5}$ (0.4). Detections of such high LAE number densities at $z \sim 6$ would be a clear signature of an over-dense region that could evolve and resemble the local group volume at $z=0$. 

\end{abstract} 


\section{Introduction} 
\label{intro}
The search for high-redshift LAEs has advanced tremendously since the days of their first successful detections, e.g. Lowenthal et al. (1991). Advances in instrument sensitivity, refined selection techniques and specific LAE spectral signatures have led to the confirmed detections of hundreds of LAEs in a wide redshift range, between $z \sim 2.2-6.6$ (e.g. Cowie \& Hu 1998; Steidel et al. 2000; Malhotra et al. 2005; Shimasaku et al. 2006; Kashikawa et al. 2006; Hu et al. 2010).
 
Over the past few years, LAEs have been used extensively as probes of reionization, high redshift galaxy evolution and the dust content of early galaxies (Dijkstra et al. 2007; Kobayashi et al. 2007; Dayal et al. 2009; Finkelstein et al. 2009; Dayal, Ferrara \& Saro 2010; Dayal, Maselli \& Ferrara 2011). However, scant effort has been devoted to establishing a link between such high-redshift observations ($z \sim 6$) to those at $z = 0$, especially in the context of the local group. 

The first such attempt was made by Salvadori, Dayal \& Ferrara (2010), who coupled the semi-analytic code, {\tt GAMETE} (Salvadori, Schneider \& Ferrara 2007; Salvadori \& Ferrara 2009), which successfully reproduces most of the observed MW and dwarf satellite properties at $z=0$, to a LAE model developed in Dayal et al. (2009) and Dayal, Ferrara \& Saro (2010). Though a powerful statistical tool, the semi-analytic nature of {\tt GAMETE} meant that the effects of clustered sources on the visibility of LAEs could not be properly considered; to alleviate such a problem to some extent, all the calculations presented therein were carried out assuming the IGM to be completely ionized. 

In this work, we extend the calculations presented in Salvadori, Dayal \& Ferrara (2010). We use a $z \sim 6$ snapshot from the {\tt CLUES} runs, which are state of the art constrained simulations that build up one of the possible merger histories of the local group, including the MW and M31 galaxies. This snapshot is coupled to a LAE model presented in Dayal, Ferrara \& Saro (2010) and Dayal, Maselli \& Ferrara (2011), that reproduces a number of data sets accumulated for high-z LAEs. The main advantage of using these simulations lies in the fact that they provide information on both the physical properties and the spatial positions of all the progenitors of the local group at any given redshift. This allows us to calculate the effects of clustered sources on the visibility of such progenitors in the Ly$\alpha$, for IGM ionization states ranging from a completely neutral ($\chi_{HI}=0.99$) to a fully ionized ($\chi_{HI}=10^{-5}$) one. 

The main questions we aim to answer with such an approach are: (a) how reionized does the volume containing the local group have to be at $z \sim 6$ for any of its progenitors to be visible as a LAE?, (b) where do such LAEs lie with respect to the underlying dark matter (DM) distribution?, and (c) what are the physical properties of the progenitors that are visible as LAEs, and how do they compare to those of field LAEs observed in cosmological searches at $z \sim 6$, e.g. by Shimasaku et al. (2006).
\section{The CLUES Simulations}
\label{clues}
For the calculations presented in this \textit{Letter}, we use the {\tt CLUES} simulations (http://www.clues-project.org). We briefly summarize the simulations here and the interested reader is referred to
Libeskind et al. (2010), Knebe et al. (2010), Libeskind et al. (2011) and
Knebe et al. (2011) for complete details on the central galaxies,
their discs and the properties of their $z=0$ substructure population,
as well as details regarding the simulation initial conditions, gas-dynamics and
star formation. 

The simulations used in this study were run with the PMTree-SPH MPI code \texttt{GADGET2}
(Springel 2005) in a cosmological box of size $64 h^{-1}$ comoving Mpc ($\rm {cMpc}$).
The runs used standard $\Lambda$CDM initial conditions and WMAP3 parameters
(Spergel et al. 2007) such that $\Omega_m = 0.24$, $\Omega_{b} =
0.042$, $\Omega_{\Lambda} = 0.76$, $h = 0.73$, $\sigma_8 = 0.73$ and
$n=0.95$. The initial conditions were constrained using observations of
peculiar velocities and the positions of objects in the local volume
(Hoffman \& Ribak 1991). Since this method only constrains large (i.e.
linear) scale structures, a local group (defined in terms of the mass,
relative distance and number of members) was selected from a number of
constrained low resolution simulations. 

\begin{figure*} 
  \vspace*{10pt} 
\center{\includegraphics[scale=0.85]{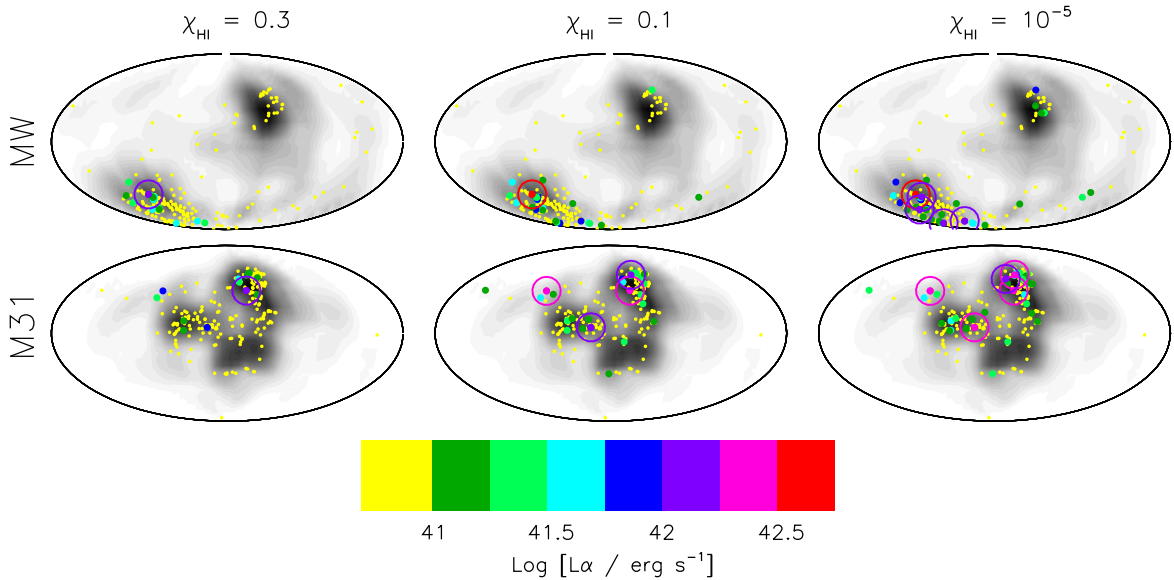}} 
\caption{Sky projections of the progenitors of the MW (upper panel) and M31 (lower panel) at $z \sim 6$ in a radius of $2 h^{-1} {\rm cMpc}$. Points and circles show the Ly$\alpha$ luminosity of the progenitors of the local group, and the LAE$_{LG}$ respectively, for the initial $\chi_{HI}$ values marked over each projection. The luminosity is color-coded according to the bar shown between $10^{41-43}{\rm erg \, s^{-1}}$ and all progenitors with $L_\alpha<10^{41} {\rm erg \, s^{-1}}$ are shown with yellow points. While the LAE$_{LG}$ are detectable with current instrument capabilities, all progenitors with $L_\alpha> 10^{41} {\rm erg \, s^{-1}}$ are within the sensitivity of future instruments, e.g. the Multi Unit Spectroscopic Explorer (MUSE). The DM density field in the $2 h^{-1} {\rm cMpc}$ box is shown by contours such that the density is highest in the darkest shaded regions, and reaches the average cosmological value for the regions colored white.  }
\label{skyproj} 
\end{figure*} 

The local group, of volume $2h^{-1}{\rm cMpc}$ at $z=0$, was re-simulated at a much
higher resolution using the prescriptions given by Klypin et al. (2001), to produce two objects resembling the MW and M31,
each of which contains about $10^6$ particles within their virial radii. Outside of this region, the simulation box was populated with lower resolution
(i.e higher mass) particles to mimic the correct surrounding
environment. In the re-simulated region, the DM and gas particle masses were taken to be $2.1\times 10^5 h^{-1}{\rm M_\odot}$ and $4.4 \times
10^4 h^{-1} {\rm M_\odot}$, respectively. The initial mass function
used is Salpeter between $0.1-100{\rm M_\odot}$. We used the star
formation rules of {\tt GADGET2} with minor modifications (see
Libeskind et al. 2010 for details); each gas particle was allowed to
undergo two episodes of star formation, each time spawning a star
particle of half the original mass (i.e. $2.2\times10^{4}~h^{-1}{\rm
M_{\odot}}$). After the simulation was run to $z=0$, we identified
cosmological structures using the MPI+OpenMP Amiga halo finder (AHF;
Knollmann \& Knebe 2009) in each snapshot.

In order to find the local group LAEs at $z\sim 6$, we
identify all the particles within the virial radius of MW and M31 at
$z=0$. These particles are then followed back in time and located at
$z\sim 6$, in the simulation box. If bound to a structure, we identify the corresponding halo: each of these
halos was then considered a progenitor of the local group at $z \sim 6$
and used in the LAE calculations, as explained in what follows.

\section{Lyman alpha visibility}
\label{find_laes}
We start the calculations by obtaining the mass, metallicity and redshift of formation of each star particle in each of the MW/M31 progenitors, in addition to the global physical properties (total halo/stellar/gas mass) of each progenitor; the age for each star particle is calculated as the time difference between its formation redshift, $z_{form}$ and $z \sim 6$. We find 234 and 240 star forming progenitor haloes of the MW and M31 respectively. Considering that stars form in a burst after which they evolve passively, we use the population synthesis code {\tt STARBURST99} (Leitherer et al. 1999) to obtain the intrinsic spectrum for each star particle depending on its age, mass and metallicity; the latter two values refer to the mass and metallicity of the star particle at the time of its formation. The total intrinsic Ly$\alpha$/continuum luminosity (1216 and 1375 \AA\, in the galaxy rest frame respectively) produced by stellar sources for each progenitor is then the sum of the Ly$\alpha$/continuum luminosity produced by all its star particles. The intrinsic Ly$\alpha$ luminosity can be translated into the observed luminosity such that $L_\alpha = L_\alpha^{int} f_\alpha T_\alpha$, while the observed continuum luminosity, $L_c$ is expressed as $L_c = L_c^{int} f_c$. Here, $f_\alpha$ ($f_c$) are the fractions of Ly$\alpha$ (continuum) photons escaping the galactic environment and $T_\alpha$ is the fraction of the Ly$\alpha$ photons that are transmitted through the IGM. 

We start by summarizing the calculation of $f_c$: for each progenitor, the dust enrichment is calculated assuming Type II supernovae (SNII) to be the primary dust factories (Todini \& Ferrara 2001). We assume that each SNII produces about $0.5 M_\odot$ of dust (Todini \& Ferrara 2001), each SNII destroys dust with an efficiency of about 40\% the region it shocks to speeds $\geq 100~{\rm km \, s^{-1}}$ (e.g. Seab \& Shull 1983), a homogeneous mixture of gas and dust is astrated into star formation, and that dust is lost in SNII powered outflows. We assume a slab-like dust distribution such that the dust distribution scale, $r_d\sim 0.5 r_g$, where the gas distribution radius is calculated as $r_g=4.5\lambda r_{200}$; the spin parameter is taken to be $\lambda=0.05$ (Ferrara, Pettini \& Shchekinov 2000) and $r_{200}$ is the virial radius. We then use $f_\alpha = 1.5 f_c$, as inferred for LAEs at $z \approx 6$ (see Dayal, Ferrara \& Saro 2010 for complete details of this calculation).

Ly$\alpha$ photons are further attenuated by the \HI present in the IGM and only a fraction $0 < T_\alpha < 1$ reach the observer. Since the IGM ionization state is largely unconstrained at $z \sim 6$, we explore 14 different values ranging from an almost neutral to completely ionized IGM such that $\chi_{HI}=0.99, 0.9, 0.8, 0.7, 0.6, 0.5, 0.4, 0.3, 0.2,0.1, 10^{-2}, 10^{-3}, 10^{-4}$, $10^{-5}$. For each such initial value of $\chi_{HI}$ (i.e. the value before the effects of the ionization regions built by each progenitor are considered), we start by computing the radius of the \HII region each progenitor ionizes around itself; we use a \HI ionizing photon escape fraction of 2\%, following the results obtained by Gnedin et al. (2008). However, in reality, multiple galaxies generally contribute ionizing photons to the same ionized region due to source clustering, which is then characterized by an effective radius. Then, within the effective ionized region each galaxy is embedded in, the total photoionization rate seen by it includes the direct radiation from the galaxy itself, from the galaxies clustered around it, and from the ultraviolet background. This procedure is carried out for each galaxy in the simulated volume. Assuming photoionization equilibrium within the effective ionized region of each galaxy and forcing $\chi_{HI}$ to attain the assigned global value at the edge of this region, we use the Voigt profile to calculate the optical depth, and hence $T_\alpha$ for Ly$\alpha$ photons along the line of sight. Complete details of this calculation can be found in Dayal \& Ferrara (2011).

Progenitors of the local group are then identified as LAEs based on the currently used observational criterion: $L_\alpha \geq 10^{42} \, {\rm erg \, s^{-1}}$ and the observed equivalent width, $L_\alpha/L_c \geq 20$~\AA. Progenitors visible as LAEs for any IGM ionization state are then referred to as LAE$_{LG}$.  

\section{Results}
\label{results}
We now answer the first question we posed: how reionized does the local group volume have to be at $z \sim 6$ for any of its progenitors to be visible as LAEs? We find that including the effects of clustered sources, the first LAE$_{LG}$ (which is a progenitor of the MW) appears when the average \HI fraction is $\chi_{HI}=0.4$; if clustering is not considered, the first LAE$_{LG}$ appears as late as $\chi_{HI}=0.1$. Progressively lower values of $\chi_{HI}$ lead to more progenitors becoming visible in the Ly$\alpha$ as expected: 1 progenitor each of the MW/M31 is visible for $\chi_{HI}=0.3$, 1 (4) progenitors of the MW (M31) are visible as LAE$_{LG}$ for $\chi_{HI} =0.1$, while the number rises to 5 each for the MW and M31 for $\chi_{HI} = 10^{-5}$, as seen from the left-most to right-most projections for the MW (M31) in the top (bottom) panels of Fig. 1. These results can be explained as follows: for a given stellar mass (or star formation rate) and age, the size of the \HII region that any progenitor can ionize around itself increases with decreasing values of $\chi_{HI}$; a lower value of $\chi_{HI}$ in the IGM also results in a larger number of clusterers. Both these effects lead to a lower value of $\chi_{HI}$ at each point within the \HII region, which itself is now larger. As a result, the Ly$\alpha$ photons face a lower optical depth at each point within this ionized region and are more red-shifted when they reach the edge of such a region; these two effects then boost $T_\alpha$. As expected, all the LAEs become progressively brighter in the Ly$\alpha$ as the value of $\chi_{HI}$ decreases, as seen from a comparison of the observed Ly$\alpha$ luminosity plotted (panels left to right), for decreasing $\chi_{HI}$ for both the MW and M31, as shown in Fig. 1. As for the distribution of the LAE$_{LG}$ with respect to the underlying DM density field, we find that for any value of $\chi_{HI}$, LAEs lie in dense DM filaments where the halo masses are large enough so that the gas in them can cool and form stars, as shown in Fig. 1. 

\begin{figure*} 
  \vspace*{10pt} 
\center{\includegraphics[scale=0.8]{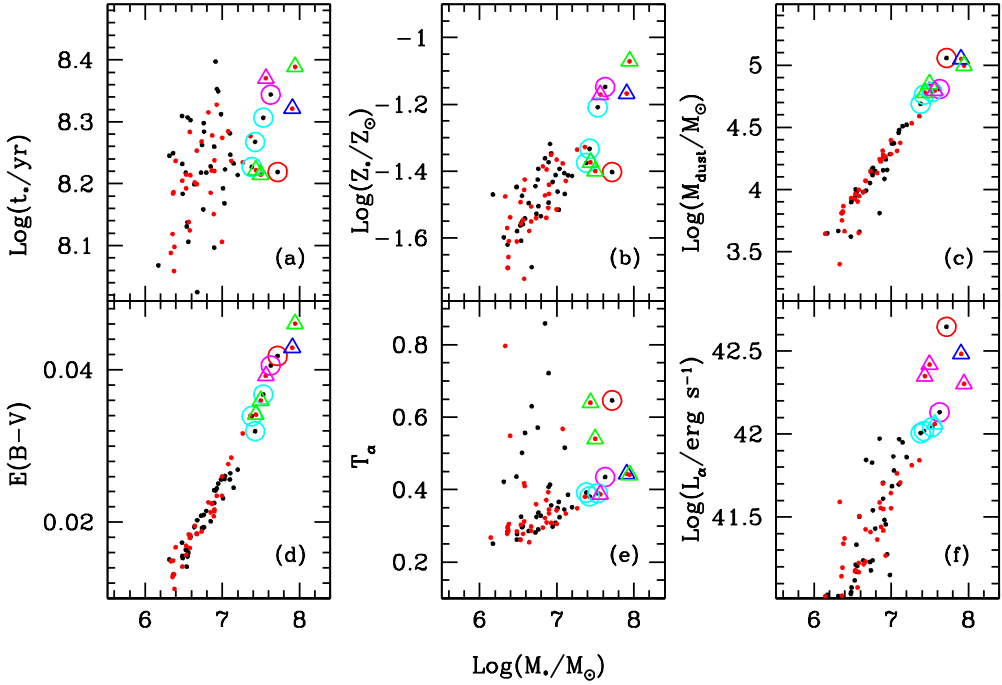}} 
\caption{Summary of the physical properties of the local group progenitors at $z\sim 6$. In each Panel, black and red points represent the progenitors of the MW and M31 obtained from the {\tt CLUES} runs. The LAE$_{LG}$ of the MW and M31 are shown by colored circular and triangular symbols, respectively. The different colors show the progenitors that become visible as LAEs for different initial IGM ionization states: $\chi_{HI}=0.4$ (red), $0.3$ (blue), $0.2$ (green), $10^{-2}$ (magneta) and $10^{-5}$ (cyan). As a function of the total stellar mass ($M_*$), the panels represent (a) the stellar mass weighted age ($t_*$), (b) the mass weighted stellar metallicity ($Z_*$), (c) the total dust mass ($M_{dust}$), (d) the color excess $E(B-V)$, (e) fraction of Ly$\alpha$ luminosity transmitted through the IGM ($T_\alpha$) and (f) the observed Ly$\alpha$ luminosity ($L_\alpha$). The last two values are plotted for $\chi_{HI} \sim 10^{-5}$; these then represent the upper bounds on the quantities obtained from our model.}
\label{phy_prop} 
\end{figure*} 

As for the physical properties of the LAE$_{LG}$, we start with the total stellar mass, which is possibly the best constrained physical quantity, when FIR data is available (e.g. from Spitzer at $3.6, 4.5\mu m$). The LAE$_{LG}$ are amongst the most massive progenitors and have stellar mass $M_* \sim 10^{7.2-8} {\rm M_\odot}$; for progenitors with lower masses, a combination of dust and IGM transmission reduces $L_\alpha$ below $10^{42} {\rm erg\, s^{-1}}$. As expected in a hierarchical structure formation scenario, LAE$_{LG}$ with the largest stellar mass have taken the longest to assemble and have stellar mass weighted ages $t_* \sim 150-250$ Myr as shown in panel (a) of Fig. 2, implying that many of these progenitors started forming stars at $z \geq 7.6$. However, we stress that it is the young stars that formed over the last $\sim 30$ Myr that dominate the stellar luminosity in the entire wavelength range from the rest-frame UV to the near Infrared. It is interesting to see that it is not necessarily the progenitors with the largest stellar mass that become visible for the largest $\chi_{HI}$ since the visibility sensitively depends on the contribution due to clustered sources; such contribution does not correlate with the progenitor mass, as is explained later. 

As star formation is the main source of metals, progenitors that are able to build up the largest stellar mass are also the most metal rich, as seen from panel (b) of Fig. 2, where the mass weighted stellar metallicity for the LAE$_{LG}$ ranges between $Z_* = 0.04-0.1 {\rm Z_\odot}$. Since the dust mass is calculated using the SFR averaged over the entire star formation history of each progenitor ($=M_*/t_*$), and since all the LAE$_{LG}$ have ages between $t_*=150-250$ Myr as mentioned, progenitors with the largest stellar mass are also the most dust rich with dust mass $M_{dust} \sim 10^{4.6-5.1} {\rm M_\odot}$, as seen from panel (c) of Fig. 2. To translate the dust mass into the color excess, we use the supernova extinction curve (Bianchi \& Schneider 2007). The color excess is then related to the escape fraction of continuum photons as $E(B-V) = -2.5 {\rm log}10 (f_c)/ 11.08$, using which $E(B-V) = 0.03-0.048$ for the LAE$_{LG}$ as shown in panel (d) of Fig. 2; the results quoted here do not change using the Calzetti extinction curve where $E(B-V) = -2.5 {\rm log}10 (f_c)/ 10.9$. 

As for the IGM transmission for LAE$_{LG}$, it shows a scatter between $T_\alpha \sim 0.35-0.65$ due to the varying contributions to the photoionization rates of the LAE$_{LG}$, from clustered sources. This clearly highlights the importance of such sources even for values of $\chi_{HI}$ as low as $10^{-5}$, as seen from panel (e) of Fig. 2; it is not necessarily the largest galaxies that have the largest contribution from clustering, as seen from the same panel. Finally, we show the observed Ly$\alpha$ luminosity for the LAE$_{LG}$ for $\chi_{HI} = 10^{-5}$: the scatter in $T_\alpha$ translates in a scatter into the observed Ly$\alpha$ luminosity, which ranges between $10^{42-42.7} {\rm erg\, s^{-1}}$ as shown in panel (f) of the same figure. As the value of $\chi_{HI}$ increases, both the value of $T_\alpha$ and $L_\alpha$ decrease. Therefore, the values shown in panels (e) and (f) represent the upper limits on the quantities shown. We can now compare the properties of the LAE$_{LG}$ to those of the field LAEs in cosmological volumes ($\sim 10^6 {\rm cMpc}^3)$. The LAE$_{LG}$ lie at the lower mass end of the general LAE population, which have stellar masses $\sim 10^{8-10.5} {\rm M_\odot}$ (panel a1 of Fig. 6, Dayal et al. 2009), stellar metallicity between $0.01-0.5 {\rm Z_\odot}$ (panel a3 of Fig. 6, Dayal et al. 2009), dust mass between $10^{4-7.2} {\rm M_\odot}$ (Panel a of Fig. 5, Dayal, Ferrara \& Saro 2010) and an average color excess $E(B-V) \sim 0.14$ (Dayal, Ferrara \& Saro 2010). 

Finally, we find that in our re-simulated volume ($\sim 86 \,{\rm cMpc^3}$) the number density of LAE$_{LG}$ is $10^{-0.93}\, {\rm cMpc}^{-3}$ for $\chi_{HI} = 10^{-5}$. For a similar value of $\chi_{HI}$, we sample a cosmological simulation snapshot at $z \sim 6$ (described in Dayal, Ferrara \& Saro 2010, which has a volume of $10^6 {\rm cMpc^3}$), about ten thousand times by randomly placing the re-simulated volume within it. We find that the number density of field LAEs is about two orders of magnitude lower, $\sim 10^{-2.86}\, {\rm cMpc}^{-3}$. Even for $\chi_{HI}=0.4$, the number density of LAE$_{LG}$ ($10^{-1.93}\, {\rm cMpc}^{-3}$) is higher than that of the field LAEs by about one order of magnitude. Detection of such a {\it large over-density} of LAEs in a volume similar to the re-simulated one could be an excellent signature of a region that could evolve to resemble the local group volume at $z \sim 0$. However, we caution the reader that to what extent such a result depends on the different resolutions of these two simulations still needs to be explored in full detail.

\section{Conclusions and discussion}
\label{conc}
We present the first work at high-z ($z \sim 6$) that couples state of the art high resolution gas-dynamical simulations of the local group run within the \texttt{CLUES} framework, with a LAE model to identify the local group progenitors (designated LAE$_{LG}$) that could be visible as LAEs at a time when the Universe was only about 1 Gyr old. The main results from this study are now summarized:

\begin{itemize}

\item The first LAE$_{LG}$, which is a progenitor of the MW, appears at an average IGM neutral hydrogen fraction of $\chi_{HI}=0.4$ (0.1) including (neglecting) the effects of clustered sources. As $\chi_{HI}$ decreases to $10^{-5}$, 5 progenitors each of the MW and M31 become visible as LAE$_{LG}$. 

\item $T_\alpha$ shows a large scatter, ranging between $0.35-0.65$ at $\chi_{HI}=10^{-5}$, due to the varying photoionization rate contributions from the clustered sources of the LAE$_{LG}$; clustering is imperative in making many progenitors visible in the Ly$\alpha$ even for a fully ionized IGM.

\item The LAE$_{LG}$ lie at the low mass end of the field LAEs: they have $M_* \sim 10^{7.2-8} {\rm M_\odot}$, $M_{dust} \sim 10^{4.6-5.1} {\rm M_\odot}$ and $E(B-V) \sim 0.03-0.048$. On the other hand, field LAEs are much larger and have $M_* \sim 10^{8-10.5} {\rm M_\odot}$, $M_{dust} \sim 10^{4-7.2} {\rm M_\odot}$ and an average $E(B-V) =0.14$. 

\item Finally, our results suggest an observable imprint of such high-z LAE$_{LG}$: the number density of LAE$_{LG}$ is about two (one) orders of magnitude higher than that of field LAEs for $\chi_{HI} = 10^{-5}$ ($\chi_{HI}=0.4$). Such {\it high number-densities} at $z \sim 6$ would be excellent signatures of a region that could resemble the local group volume at $z \sim 0$. However, the extent to which such a result depends on the varying simulation resolutions still needs to be explored in more detail.

\end{itemize}

The main caveats involved in this study are with regards to the distribution of dust (homogeneous/clumpy) in the interstellar medium of these high-z progenitors (e.g. Finkelstein et al. 2009; Dayal, Ferrara \& Saro 2010; Dayal, Maselli \& Ferrara 2011), and the effects of peculiar velocities (inflows/outflows into/from the galaxies) that can have enormous impact on $T_\alpha$ (Verhamme et al. 2006; Dayal, Maselli \& Ferrara 2011). Uncertainties also remain on the value of the escape fraction of \HI ionizing photons used. As discussed in the above mentioned works, it is unclear to what extent these effects can modify the observed Ly$\alpha$ luminosity. It is hoped that a combination of the expected data from upcoming missions such as ALMA, MUSE and JWST will be fundamental in resolving such issues.


\section{Acknowledgements}
PD thanks SISSA for their generous allocation of cluster time. NIL is supported through a grant from the Deutsche Forschungs Gemeinschaft. The simulations were performed and analyzed at the Leibniz Rechenzentrum Munich (LRZ), the Neumann Institute for Computing (NIC) Juelich and at the Barcelona Supercomputing Centre (BSC). The authors thank A. Ferrara, S. Gottl\"ober, S. Nuza, S. Salvadori and the referee for positive comments. 


\label{lastpage} 
\end{document}